\documentclass[12pt]{article}
\usepackage{citesort}
\usepackage{latexsym} 
\newcommand{\beq}{\begin{equation}}
\newcommand{\eeq}{\end{equation}}
\newcommand{\bea}{\begin{eqnarray}}
\newcommand{\eea}{\end{eqnarray}}
\newcommand{\cj}{{\cal J}}

\begin{document}

\begin{titlepage}
\begin{flushright}
{\small SAGA-HE-219 \\ KYUSHU-HET-82} \\ hep-ph/0503017
\end{flushright}
\vspace*{3mm}

\begin{center}
{\large\bf 
Localized Kaluza-Klein Graviton and Cosmological Constant}
\vspace*{12mm}

Naohiro Muta\footnote{muta@dirac.phys.saga-u.ac.jp} 
and Nobuhiro Uekusa\footnote{uekusa@higgs.phys.kyushu-u.ac.jp}\\
\vspace*{4mm}

${}^{*}\!\!${\textit{
Department of Physics, Saga University, Saga 840-8502, Japan}}\\
${}^{\dag}\!\!${\textit{
Department of Physics, Kyushu University, Fukuoka 812-8581, Japan}}
\end{center}
\vspace*{10mm}

\begin{abstract}
\vspace{.3cm}

We study linearized graviton in the presence of 
a four-dimensional cosmological constant
in two brane models with a warped extra dimension.
In explicit models including bulk scalar fields,
we calculate
the masses of Kaluza-Klein modes of graviton
and their interactions with matter on the visible brane.
It is shown that the effects of the cosmological constant 
contribute by the equivalent size to the warp factor, masses and couplings
and that bulk scalar fields can increase the effects.
This is examined 
further independently of the forms of scalar potentials.
Then it is found
how the masses and couplings are described in terms of 
the warp factors and generic scalar potentials.
A possibility that the masses and couplings are significantly 
changed by cosmological constant effects
is discussed.

\end{abstract}
\end{titlepage}

\newpage
\renewcommand{\theequation}{\thesection.\arabic{equation}} 
\setcounter{equation}{0}
\section{Introduction}

There has been much interest in phenomenological possibilities of
higher-dimensional models with a warped extra dimension since 
it was shown that they may provide a solution to the gauge hierarchy
problem.
In the Randall-Sundrum model~\cite{RS},
the first Kaluza-Klein (KK) mode of graviton can have
the mass of the order ${\cal O}(1)$ TeV and 
the coupling with matter on the visible brane is
of the order ${\cal O}(1)$ TeV${}^{-1}$.
This induces new effects which in principle can be seen
at future colliders.
The first KK graviton will be directly searched in the resonance 
production such as 
Drell-Yan process~\cite{DHR}
or electron-positron pair annihilation~\cite{elle}.
Models with a warped extra dimension extensively has been studied  
also in phenomenological context including
brane localized curvature, radion,
bulk gauge bosons, bulk fermions,
neutrino oscillations,
grand unification,
and supersymmetry.
The progress has led to various possible ways of 
detecting an extra dimension. 
It however has been assumed that in most of these models 
the fine-tuning between the bulk curvature 
and brane tension makes
the four-dimensional cosmological constant vanish.
Since recent observations have supported that 
the cosmological constant is nonzero~\cite{observ},
it is interesting to examine
new phenomenological possibilities at TeV scale
by taking in this effect.

In the nonzero cosmological constant case,
brane picture including localizability and
stability has been inspected.
Brane configurations with a cosmological constant are solutions to 
the setup in the Randall-Sundrum model, where the fine-tuning
is relaxed~\cite{kr}. 
If bulk scalar fields are taken into account,
it also may be possible to address a explanation to
the smallness of the cosmological constant 
with less tuning~\cite{self,cf,wph,wphex}.
For these solutions, it has been found that
localization of fields~\cite{locc}
and Newton's potential~\cite{newc} are analogous
to those of the zero cosmological constant case~\cite{RS,loc,new}
and that radion with a stable mass spectrum 
seems to need bulk scalar fields~\cite{ccr}.
Since bulk scalar fields change background geometry,
it is likely that they play 
a role to change phenomenological consequences.

In this paper, we calculate the masses and interactions of KK graviton 
on the visible brane in models including
a warped extra dimension,
a four-dimensional cosmological constant and
bulk scalar fields.
We first perform the analyses
in explicit models where the warp factors are expanded 
into power series of the cosmological constant.
One of the models is the pure gravity case and 
the other is the case of the scalar potential given 
in Ref.~\cite{cf}. 
From the analyses at the first order of the power expansion,
it is found that 
the masses and couplings receive
more largely the effects of the cosmological constant
in the model with the bulk scalar fields.
This is understood from that the bulk scalar field increases
the cosmological constant effect on the warp factor.
Beyond the power expansion, 
it is required to examine whether bulk scalar fields
enhance the cosmological constant effects. 
We analyze mode eigenfunction of graviton
for generic warp factors
without relying on 
the power expansion and special forms of scalar potentials.
It is found how the masses and couplings are generally
described by the warp factors and scalar potentials.
Then it is shown that masses and couplings
significantly can be different from those of the zero cosmological constant 
case.

This paper is organized as follows.
In Section 2, 
we briefly review the formulation of
KK modes of graviton in models with a warped extra dimension
and a four-dimensional cosmological constant.
Section 3 are devoted to analyzing masses and interactions
in explicit models.
A relation of the warp factors 
with the masses and couplings is found.
In Section 4, 
we derive a more general relation 
among warp factors, masses, and couplings.
We summarize our results in Section 5.

\setcounter{equation}{0}
\section{Formulation}

The model is formulated in the five-dimensional general relativity
on the orbifold S$^1$/Z$_2$ with the compactification radius $R$.
The action of the model is
\beq
  S=
  {1\over 2\kappa^2}
  \int d^4 x \int dy \sqrt{-\det g_{MN}}\left({\cal R}
 -\partial_M \Phi \partial^M \Phi- 2V(\Phi) 
  \right) 
  +S_{\textrm{\scriptsize source}}+S_{\textrm{\scriptsize matter}},
 \label{oact}
\eeq
with the five-dimensional gravitational coupling constant $\kappa$, 
the five-dimensional Ricci scalar ${\cal R}$
and the scalar potential $V$. 
The four-dimensional and fifth space coordinates
are denoted by $x$ and $y$, respectively,
and the capital Latin indices $M$, $N$, $\dots$ label
the five-dimensional indices.
The action $S_{\textrm{\scriptsize source}}$ is
introduced as a source in order that two branes are consistently set
at the fixed points of the orbifold.
This action plays roles to parameterize the tensions of the branes and
to stabilize the compactification radius. 
We simply assume that the source action controls   
the singularities along the direction perpendicular to the branes
and that the cutoff energies on the branes are 
Planck scale at $y=0$ and TeV scale at $y=\pi R$.
In the equation (\ref{oact}), the last term $S_{\textrm{\scriptsize matter}}$
describes fields in the visible sector,
which are confined to the brane at $y=\pi R$.
The interactions of the bulk fields with these fields induce new effects 
in scattering processes at TeV scale.

We work with the scalar background independent of the
coordinates parallel to the branes, $\Phi=\Phi(y)$ 
and the line element
\beq
ds^2=A(y)^2(-dt^2+e^{2\sqrt{\lambda}t}~ \delta_{ij}dx^i dx^j) +dy^2  ,
\label{bg}
\eeq
where $\lambda$ indicates
the four-dimensional cosmological constant.
Although $\lambda$ is not a parameter of the model,
the smallness can be chosen by hand with the degrees of freedom of
parameterizing the brane tension\footnote{The warp factor may be used 
for the purpose of explaining 
the small cosmological constant
instead of the gauge hierarchy problem~\cite{self,cf,wph,wphex}.
We will not discuss the matter further in this paper.}.
The warp factor $A$ necessarily involves $\lambda$
dependently on the scalar background
since the line element is the solution to the equations of motion derived 
from the action (\ref{oact}).
In the equations of motion,
the scalar potential can be composed of the superpotential-like
function $W(\Phi)$,
\beq
   V(\Phi)={1\over 8\gamma^2}\left({\partial W\over \partial \Phi}\right)^2
      -{1\over 6}W^2 ,
  \label{poten}
\eeq
where $\gamma(y)=\sqrt{1+4\lambda/ (WA)^2}$.
Then the other equations are written as
the first order equations  
\beq
   {A'\over A}=-{1\over 6}W\gamma ,~~~~~
    \Phi'={1\over 2\gamma}{\partial W\over \partial\Phi} ,
  \label{1st}
\eeq
where the prime denotes the derivative with respect to $y$.
Explicit forms of $W$ are given in Section~\ref{expmc}.
In the following, we describe fluctuations.
We concentrate our attention on analyses of the tensor modes
of graviton,
since the fluctuations decouple from 
the scalar and vector mode fluctuations
and cannot be removed by any gauge choice~\cite{gf}.

The tensor mode fluctuations are taken as
the line element (\ref{bg}) with the replacement 
$\delta_{ij}\rightarrow \delta_{ij}+\kappa h_{ij}(x,y)$.
The fluctuations are decomposed over eigenfunctions:
\beq  
 h_{ij}(x,y)={1\over \sqrt{R}}\left[h_{ij}^{(0)}(x)
   +\sum_{n=1}^{\infty}h_{ij}^{(n)}(x)\, \chi^{(n)}(y)\right] ,
  \label{kkdec}
\eeq
where $h_{ij}^{(0)}$ and $h_{ij}^{(n)}$ stand for
the zero mode and the $n$-th KK mode respectively.
The eigenfunctions $\chi^{(n)}$ are determined such that 
the four-dimensional spectra are governed by ordinary four-dimensional
equations of motion.
From the action (\ref{oact}) and the replaced line element,
the kinetic action quadratic of tensor fluctuations are 
\bea
S_{\textrm{\scriptsize kin}}
&\!\!\!=\!\!\!&
 \int_0^{\pi R} {dy\over R}  A^2
  \int d^4x \sqrt{-\widetilde{g}}~ h_{ij}^{(0)}\Box h_{ij}^{(0)}
\nonumber
\\     
 &&
+\sum_{n=1}^{\infty}\int_0^{\pi R}{dy\over R} A^2 \chi^{(n)}{}^2
 \int d^4 x\sqrt{-\widetilde{g}}~ h_{ij}^{(n)}
\left(\Box-m_n^2\right) h_{ij}^{(n)} ,
\label{tenac}
\eea
up to numerical factors.
The determinant $\widetilde{g}$ and the d'Alembertian $\Box$ 
are made out of the four-dimensional metric
$d\widetilde{s}^{2}=-dt^2+e^{2\sqrt{\lambda}t}\,\delta_{ij}dx^i dx^j$.
The zero mode is massless and the KK modes has the masses $m_n$.
The normalization is canonically given by
\beq
   \int_0^{\pi R}{dy\over R} A^2 \chi^{(n)}{}^2=1 .
  \label{kknorm}
\eeq
The $n$-th eigenfunction $\chi^{(n)}$ must satisfy the equation
\beq
\left(\partial_y^2 +4{A'\over A}\partial_y 
 +{m_n^2\over A^2}\right)\chi^{(n)}(y) = 0 ,
\label{deqm}
\eeq
which automatically leads to the orthogonality of the eigenfunctions.
The boundary conditions are
\beq
\partial_y \chi^{(n)}(y)=0 , \quad \textrm{at}\quad y=0,~ \pi R .
\label{bc0}
\eeq
It is seen that the $n$-th KK mass is derived from
the equations (\ref{deqm}) and (\ref{bc0}), as follows.
In the equation (\ref{deqm}),
the eigenfunction $\chi^{(n)}$ is solved with
two integration constants, one of which 
is an overall constant.
In the solution,
the other integration constant and the mass $m_n$
are determined by
the two boundary conditions (\ref{bc0}).
Thus the mass is completely described
by the parameters of the model.

The interaction of the tensor fluctuations with matter is given by
\beq
  S_{\textrm{\scriptsize int}}
  ={\kappa\over \sqrt{R}}\int d^4x \sqrt{-g_4}\,h_{ij}^{(0)}T^{ij} 
   +
   \left. \sum_{n=1}^{\infty}  
   {\kappa\over \sqrt{R}}\chi^{(n)}\right|_{y=\pi R}
     \int d^4x \sqrt{-g_4}\,h_{ij}^{(n)}T^{ij} ,
 \label{kktcou}
\eeq
with the four-dimensional energy momentum tensor $T_{ij}$.
While the coupling for the zero mode is suppressed 
by $\sqrt{R}/\kappa$ which is effectively of the order of Planck scale,
the coupling for the KK modes may become large dependently on
$\chi^{(n)}$.
The size of $\chi^{(n)}$ can be obtained after
in the integral (\ref{kknorm}) 
the overall constant is fixed.

Here we comment on the bulk scalar field.
In the equations (\ref{tenac})-(\ref{kktcou}),
the scalar $\Phi$ did not appear explicitly
since the background scalar is the solution of the model
and scalar fluctuations is dropped in the linearized 
analysis.
The effects of the scalar field
are included implicitly in the warp factor.  
The distinctive scalar function $W$ would induce
the distinctive warp factor.
As shown in the following section,
this brings a change to the masses and couplings.

\setcounter{equation}{0}
\section{Explicit analyses of the KK graviton masses and couplings
\label{expmc}}
In this section, we explicitly perform analytic 
calculations of masses and couplings of KK graviton
on the visible brane.

\subsection{Pure gravity}
The first example  is the pure gravity case
\beq
   W=6/L=\textrm{constant} ,
\eeq
where $L$  denotes the five-dimensional curvature radius.
In this case, the warp factor is given by~\cite{kr}
\beq
A(y)=\cosh\left({y\over L}\right)-\sqrt{1+\lambda L^2}\,
  \sinh\left({y\over L}\right) ,
\label{warp}
\eeq
with the orbifold condition imposed.
It has been shown that the differential equation for
the eigenfunctions $\chi^{(n)}$ is transformed into a hypergeometric
equation~\cite{locc}.
The normalization however has not been given in analytic methods.
In fact it seems to be difficult to 
integrate hypergeometric functions in a finite interval.
Since we have interest in the cosmological constant effects,
we would like to advance analytic approach 
in some approximation using the smallness of the cosmological constant.

We define the dimensionless parameter
\beq
\epsilon\equiv{\lambda L^2\over 4} \ll 1,
\label{apo}
\eeq
and reexamine the KK mode fluctuations for the warp factor 
\beq
   A= e^{-y/L} \left[1+\epsilon
  \left(1-e^{2y/L}\right)\right] ,  \label{azx}
\eeq
up to ${\cal O}(\epsilon^2)$\footnote{A similar 
approximation with respect to a small cosmological constant
was used for studying issues of scalar potential, higher curvature 
and brane universe~\cite{wphex,approx,hkt}.}.
Throughout this section all the equations will be 
described up to ${\cal O}(\epsilon^2)$.
For simplicity of notation, 
we use the coordinate $w=mL e^{y/L}$ and 
the indices of the modes will be omitted hereafter
as long as we do not mention.
Then the eigenvalue equation (\ref{deqm}) reduces to 
\beq
\left[\left\{1+2\epsilon \left(1-(w/w_0)^2\right)\right\}
 {\partial^2\over \partial w^2}
 -\left\{3+2\epsilon\left(3+(w/w_0)^2\right)\right\}
 {1\over w}{\partial\over \partial w}+1\right]\chi=0 ,
\label{eqza}
\eeq
where $w_0=mL$ and we used $\sqrt{\lambda}L\,e^{\pi R/L}\ll 1$. 
We find the following solution to the equation (\ref{eqza}): 
\begin{equation}
\chi =w^2 \cj_2+\epsilon\,\left[
  w^3\cj_3+{1\over 3w_0^2}\left\{48 w^3\cj_3-15w^4 \cj_4+w^5\cj_5
			   \right\}\right] ,
 \label{gensol}
\end{equation}
with
\beq
\cj_n=c_1 J_n(w)+c_2 Y_n (w) \quad \textrm{for}~~n=2,\cdots,5 ,
\eeq
where  $c_1$ and $c_2$ are integration constants.
The solution with the first Bessel functions $J_n$
is also represented in terms of 
Gauss' hypergeometric function as 
\begin{equation}
 {1\over 8}w^4 \, 
 F\left(2-\frac{w_0}{2\sqrt{2\epsilon}}\,,
      \,2+\frac{w_0}{2\sqrt{2\epsilon}}\,,
      \,3\,;\,\frac{2\epsilon}{1+2\epsilon}\left({w\over w_0}\right)^2\right) ,
\end{equation}
up to ${\cal O}(\epsilon^2)$.

From the solution (\ref{gensol}),
we calculate 
masses of KK modes 
according to the formulation in the previous section.
After the substitution of the solution (\ref{gensol})
into the boundary conditions (\ref{bc0}) and a short calculation,
it is found that the mass eigenvalue equation reduces to
\beq
 J_1(w_1)
 +{\epsilon\over 3w_0^2}\left[(12\,w_1-w_1^3)J_2(w_1)
    +9\,w_1^2 J_1(w_1)\right] \simeq 0 , \label{appf}
\eeq
where $w_1=mL\,e^{\pi R/L}$ and we used $e^{\pi R/L}\gg 1$.
From this equation, we obtain the $n$-th KK mass
\beq
m_n={x_n\over L} e^{-\pi R/L}\left[1-{\epsilon \over 3}\left(1-{12\over x_n^2}\right) e^{2\pi R/L}\right] ,
\label{eqn:KK-masses}
\eeq
where $x_n$ indicates the $n$-th zero of the Bessel function $J_1$,
for example, $x_1=3.83$, $x_2=7.02$.
In the equation (\ref{eqn:KK-masses}), the first order term of
$\epsilon$ is multiplied
by the factor $e^{2\pi R/L}$ compared to the zero-th order term.
The origin of this factor is $e^{2y/L}$ in the warp factor (\ref{azx}).
We speculate that 
if in some setting the warp factor receives  
larger contributions from the cosmological constant,
the mass may be changed significantly.
This possibility will be investigated 
in cases including bulk scalar fields.
The analysis of the couplings can be carried out 
with the solution (\ref{gensol})
and the mass (\ref{eqn:KK-masses}).
Using the effective Planck mass $M_P^2\simeq L/\kappa^2$, 
we obtain the coupling of the first KK mode with matter
\beq
    {\kappa\over \sqrt{R}}
   \chi^{(1)}(\pi R) 
 ={\sqrt{2}\over  M_P}
    e^{\pi R/L}
   \left[1+{4\epsilon \over 3}\left(1-{3\over x_1^2}\right)
  e^{2\pi R/L}\right] ,
\label{eqn:coupling-strength}
\eeq
which is of the order of TeV${}^{-1}$.
The factor $e^{2\pi R/L}$ appears in similar to that of the mass.
The effects of the cosmological constant on the
mass and coupling
is relevant equivalently to the effect on the  warp factor.
The derivation of the mass and coupling
without requiring $e^{\pi R/L}\gg 1$
will be given in Section~\ref{srel}. 

\subsection{Scalar-gravity
\label{bulksca}}

The second explicit model is 
the case with the scalar function
\beq
W (\Phi)={6\over L}-b\Phi^2 +\epsilon \delta W ,
\eeq
with the correction $\delta W$ $(<\epsilon^{-1})$
and the parameter $b$ $(>0)$.
This form of the function $W$ has been studied in Ref.~\cite{cf}.
In this case, the warp factor is obtained as\footnote{
In Ref.~\cite{cf}, 
the solution for $A$ was not shown explicitly.}
\beq
   \ln A =-{y\over L}
 +\epsilon\Bigg[e^{4y/L}-e^{2y/L}-{1\over bL}(1-e^{-2by})\Bigg] .
\eeq
In order to make arguments clear, 
we focus on terms relevant  
near the boundaries.
Then the warp factor can be approximated effectively as
\beq
A(y) \simeq e^{-y/L}\left[1+\epsilon\left(e^{4y/L}-e^{2y/L}\right)\right] .
\label{10}
\eeq
The factor $e^{4y/L}$ is an effect induced by the mixing between
the graviton and bulk scalar field.
If other scalar functions are set,
the scalar backgrounds may change $\epsilon$ term in the warp factor.
A more general warp factor and the corresponding eigenfunctions 
are given in Appendix~\ref{gwa}.
In the warp factor (\ref{10}),
the solution for the eigenfunction is
\bea
   && \chi=w^2 \cj_2+\epsilon \Bigg[
    {1\over 3w_0^2}
    \left\{w^5 \cj_5
        -15 w^4 \cj_4
         +48 w^3 \cj_3\right\}
\nonumber
\\
  && \qquad\qquad
   +{1\over 5w_0^4}\{
     w^7\cj_7-30w^6 \cj_6
    +240w^5\cj_5-480w^4\cj_4 \} 
  \Bigg] .
  \label{sol4}
\eea
This solution is the equation (\ref{soleven})
with the nonzero coefficients  
$d_4=1$ and  $d_2=-1$.
According to analyses in the previous subsection,
we obtain the mass of the $n$-th KK mode,
\beq
    m_n={x_n\over L}e^{-\pi R/L}
 \left[1+{\epsilon \over 5}\left(1-{24\over x_n^2}\right)
 e^{4\pi R/L}\right] ,
\label{eqn:KK-masses2}
\eeq
and the coupling,
\beq
  {\kappa \over \sqrt{R}}\chi^{(1)}(\pi R) 
 ={\sqrt{2}\over  M_P}
    e^{\pi R/L} \left[ 
   1-{4\epsilon \over 5}e^{4\pi R/L}\right] .
\eeq
It is seen that 
the factor $e^{4\pi R/L}$ occurs due to $e^{4y/L}$ in the warp factor
(\ref{10}).
Therefore  in the two models where $\epsilon$ expansion is applicable,
the exponential enhancements are equivalent 
among the warp factor, masses and couplings.
The enhanced contributions can become large dependently on the scalar function.
In the following section,
we examine how scalar potentials affect the KK graviton masses an couplings
beyond $\epsilon$ expansion.

\setcounter{equation}{0}
\section{Relation among the warp factor, 
the KK graviton masses and couplings \label{srel}}
We would like to acquire a general observation about
the relation among the effects of the cosmological constant
on the warp factor, masses, and couplings.
In order to distinguish quantities
between the zero and nonzero cosmological
constant cases,
we introduce the following representations:
\beq
     A\left[\epsilon\right],~~
     V\left[\epsilon\right],~~
     \chi\left[\epsilon\right],~~
     m\left[\epsilon\right],
  \label{ep}
\eeq
for the nonzero cosmological constant and
\beq
        A\left[0\right],~~
     V\left[0\right],~~
     \chi\left[0\right],~~
     m\left[0\right] ,
  \label{000}
\eeq
for the zero cosmological constant.
In the equations (\ref{ep}) and (\ref{000}),
it is assumed that 
each corresponding quantity have the identical sign
and that the warp factor $A\left[0\right]$ maintains 
the order of cutoff energies on the two branes.
The size of effects of cosmological constant
is evaluated as the ratio
\beq
    A_{\textrm{\scriptsize ratio}}
  ={A\left[\epsilon\right]-A\left[0\right]\over 
      A\left[0\right]} ,~~
    \chi_{\textrm{\scriptsize ratio}}
  ={\chi\left[\epsilon\right]-\chi\left[0\right]\over 
      \chi\left[0\right]} ,~~    
 m_{\textrm{\scriptsize ratio}}
  ={m\left[\epsilon\right]-m\left[0\right]\over 
      m\left[0\right]} .
 \label{ratio}
\eeq

We first investigate how $\chi_{\textrm{\scriptsize ratio}}$ 
is related to
$A_{\textrm{\scriptsize ratio}}$.
Note that the coupling is
given by $\kappa\chi(y=\pi R)/\sqrt{R}$.
The normalization (\ref{kknorm}) is
\begin{eqnarray}
&&   \int^{\pi R}_0  {dy\over R}
   A\left[\epsilon\right]^2 \chi\left[\epsilon\right]^2 =1 ,
   ~\quad\textrm{for} ~~ \chi\left[\epsilon\right] ,
\\
&&   \int^{\pi R}_0  {dy\over R}
   A\left[0\right]^2 \chi\left[0\right]^2 =1 ,
     ~\quad\textrm{for} ~~ \chi\left[0\right] . 
\end{eqnarray}
Subtracting these equations each other
reduces to
\beq
   \chi_{\textrm{\scriptsize ratio}}
   =- {A_{\textrm{\scriptsize ratio}}\over 1+A_\textrm{\scriptsize ratio}} .
  \label{ca1}
\eeq
This relation is satisfied independently of $y$.
It is seen that the coupling of KK mode with matter 
are determined only by $A_{\textrm{\scriptsize ratio}}$.
When $A_{\textrm{\scriptsize ratio}}(y=\pi R)$ is negative,
the coupling  necessarily increases.
However, the coupling of the same size 
would be obtained even in the zero cosmological
constant case
if a distinctive compactification radius is chosen.
One must take the mass into account in order 
to identify the nonzero cosmological constant effects, as follows.

We evaluate the ratio $m_{\textrm{\scriptsize ratio}}$
by examining the equation (\ref{deqm}),
\begin{eqnarray}
&&   \left[\partial_y^2-4{A\left[\epsilon\right]'\over
   A\left[\epsilon\right]}
  \partial_y
   +{m\left[\epsilon\right]^2\over A\left[\epsilon\right]^2}\right] 
  \chi\left[\epsilon\right]=0 ,
   ~\quad\textrm{for}~~ \chi\left[\epsilon\right] ,
   \label{eqce1}
\\
&&
   \left[\partial_y^2-4{A\left[0\right]'\over
   A\left[0\right]}
  \partial_y
   +{m\left[0\right]^2\over A\left[0\right]^2}\right] 
  \chi\left[0\right]=0 ,
   ~\quad\textrm{for}~~ \chi\left[0\right] .
   \label{eqce2}
\end{eqnarray}
From the equations above,
the equation (\ref{eqce1}) is written as
\beq
   {m\left[\epsilon\right]^2\over
     A\left[\epsilon\right]^2}
 -{m\left[0\right]^2\over A\left[0\right]^2}
  +\left[4{ A\left[0\right]'\over A\left[0\right]}
 -6{\chi\left[0\right]'\over
     \chi\left[0\right]}
 +6{A_{\textrm{\scriptsize ratio}}'
    \over 1+A_{\textrm{\scriptsize ratio}}}\right]
   {A_{\textrm{\scriptsize ratio}}'
   \over 1+A_{\textrm{\scriptsize ratio}}}
 -{A_{\textrm{\scriptsize ratio}}''
   \over 1+A_{\textrm{\scriptsize ratio}}}
  =0 .
  \label{eqra}
\eeq
From the boundary conditions (\ref{bc0})
and the equation (\ref{ca1}),
this equation at the boundaries
becomes
\beq
   {m\left[\epsilon\right]^2  -4\epsilon/(3L^2)\over
    A\left[\epsilon\right]^2}
  -{m\left[0\right]^2 \over A\left[0\right]^2}
  +{2\over 3}(V\left[\epsilon\right]-V\left[0\right])
 =0
\eeq
where $V$ is defined by the equation (\ref{poten}).
Then we obtain
\beq
   m_{\textrm{\scriptsize ratio}}=
    -1+ (1+A_{\textrm{\scriptsize ratio}})
   \sqrt{1-{2A\left[0\right]^2\over 3m\left[0\right]^2}
    (V\left[\epsilon\right]-V\left[0\right])
   +{4\epsilon\over 3m\left[0\right]^2 L^2 
   (1+A_{\textrm{\scriptsize ratio}})^2}} ,
   \label{mra}
\eeq
which is derived
without the simplification $e^{\pi R/L}\gg 1$.
From the assumption for $A\left[0\right]$
in the beginning of this section, 
the last term in the square root is negligible
and 
$A\left[0\right]/m\left[0\right]\sim L$.
The mass depends on the potential difference  
$(V\left[\epsilon\right]-V\left[0\right])$
as well as $A_{\textrm{\scriptsize ratio}}$.
Thus the effect of the cosmological constant is different from 
a simple choice of the compactification radius.

From the equations (\ref{ca1}) and (\ref{mra}),
the coupling and mass become of order ${\cal O}(10)$\% 
larger than those of the zero cosmological constant case
if the warp factor difference ratio 
$A_{\textrm{\scriptsize ratio}}$ 
or the scalar potential difference ratio
$(V\left[\epsilon\right]-V\left[0\right])L^2$
are a negative value of order ${\cal O}(10)$\%.
This provides a phenomenological possibility
of cosmological constant effects at TeV scale.

\setcounter{equation}{0}
\section{Summary and discussions}

We have calculated masses and interactions of KK graviton 
in models with a warped extra dimension and
a four-dimensional cosmological constant.
In the pure gravity case, 
we have reexamined the eigenfunction 
by expanding the warp factor into power series of 
the dimensionless cosmological constant $\epsilon$.
Correspondingly to the exponential factor $e^{2y/L}$
in the $\epsilon$ term of the warp factor,
it have been found that 
the mass and coupling
have the factor $e^{2\pi R/L}$ in $\epsilon$ term.
In the case with the scalar potential given in Ref.~\cite{cf},
we have found that 
the mass and coupling have the factor $e^{4\pi R/L}$ in the $\epsilon$ term
correspondingly to the warp factor $e^{4y/L}$.
Thus it has been shown that the bulk scalar field increases 
the cosmological constant effects.
In addition, the effects of
the cosmological constant on the warp factor 
are equivalent to
those of the masses and couplings. 
Beyond the power expansion,
we have also found generally 
the masses and couplings which are 
described in terms of the warp factors and scalar potentials.
From this indication,
we have presented a possibility of phenomenological effects of
the cosmological constant.

Our solutions for the graviton eigenfunctions may be useful
for studying the holographic principle~\cite{adscft,hrg}.
The original Randall-Sundrum model
is expected to be included in this context~\cite{rshol}.
It is likely that quantum theories on branes have
geometrical settings in higher-dimensional spacetime.
For quantum theories with a cosmological constant,
it was found that there is an ambiguity of the vacuum~\cite{vac}.
In fact it has been shown that this
seems to be an artifact which can be
evaded~\cite{vaci}.
It is worth while to 
seriously study quantum theories
for the nonzero cosmological constant case
and to proceed on a clear understanding of
the geometrical settings~\cite{adscftc,hrgc}.
Since our solutions for the eigenfunctions 
make the cosmological constant effects manifest,
the analyses with the solutions
would provide direct deviations from the zero cosmological constant case.

Finally we would like to mention scalar potentials.
We have not asked 
what form of scalar potential induces
phenomenologically significant cosmological constant effects.
It is interest to exhaustively investigate this question
using ansatz for scalar function.
If a viable scalar function is found,
it is also interesting to examine
how the models can be embedded
in fundamental theory such as string theory
following proposals in Ref.~\cite{embed}.

\bigskip
\subsection*{Acknowledgments}

The authors thank K.~Ghoroku for discussions.
This work is supported by the grant-in-aid for scientific research on
priority areas: ``Progress in elementary particle physics of
the 21st century through discoveries of Higgs boson and supersymmetry''
(No.~441).


\newpage
\begin{appendix}

\setcounter{equation}{0}
\section{Eigenfunctions in $\epsilon$ expansion\label{gwa}}

In this appendix, we present eigenfunctions in the case where
the warp factor has the following general form of $\epsilon$ expansion:
\beq
  A=e^{-y/L}\left[
  1+\epsilon \sum_{n=0}^{p}d_{2n}\, e^{2ny/L}
  \right] ,
\label{12}
\eeq
where the coefficients $d_{2n}$ are order of ${\cal O}(1)$.
For graviton, the equation (\ref{deqm}) is assumed,
\beq
\left(\partial_y^2+4{A'\over A}\partial_y+{m^2\over A^2}\right)\chi(y)=0 .
\label{apde}
\eeq
Then we find the following solution:
\bea
  \chi \!&=&\! w^2 \cj_{2}+\epsilon \Bigg[
  -d_{0}w^{3}\cj_{1}
  -\frac{2}{3}\frac{d_{2}}{w_0^{2}}w^{5}\cj_{1}
  +\sum_{n=0}^{1}\frac{q_{n,1}}{w_0^{2n}}{\cal I}_{n}
\nonumber
\\
  &&\qquad\qquad\qquad
  +\sum_{l=2}^{p}\left(\frac{-2}{l+2}\frac{d_{2l}}
   {w_0^{2l}}w^{2l+3}\cj_{1} 
  +\sum_{n=2}^{l}\frac{q_{n,l}}{w_0^{2n}}{\cal I}_{n}\right)
  \Bigg] ,
\label{soleven}
\eea
with 
\bea
{\cal I}_{n}
\!\!\!\!&=&\!\!\!\!%
(n+2)w^{2n+2}\cj_{2}-w^{2n+3}\cj_{1} ,
\\
q_{0,1}
\!\!\!\!&=&\!\!\!\!%
-2 d_{2} ,
\\
q_{1,1}
\!\!\!\!&=&\!\!\!\!%
-{1\over 3}d_2 ,
\\
q_{l,l}
\!\!\!\!&=&\!\!\!\!%
-{3l\over (l+2)(l+1)} d_{2l} ,\quad \textrm{for}~~ l\geq 2,
\\
q_{n,l}
\!\!\!\!&=&\!\!\!\!%
(-2)^{l-n}{l+1\over n+1}\left[ \prod_{k=n+1}^{l}{k(k-2)(k+2)\over
k+1}\right] q_{l,l} , \quad \textrm{for}~~l>n\geq 2 .
\eea

\vspace{.5cm}

Instead of the equation (\ref{apde}), we can assume the following
eigenfunction equation
for bulk gauge boson, $\chi^{(A)}$: 
\beq
\left(\partial_y^2+2{A'\over A}
\partial_y+{m^2\over A^2}\right)\chi^{(A)}(y)=0 .
\eeq
Then we obtain the following solution:
\bea
 \chi^{(A)}(y)\!&=&\! w \cj_{1}+\epsilon\Bigg[
  d_{0}(w\cj_{1}-w^{2}\cj_{0})
\nonumber
\\
 &&\qquad\qquad
 +\sum_{l=1}^{p}\left(\frac{-1}{l+1}\frac{d_{2l}}
  {w_0^{2l}}w^{2l+2}\cj_{0}
  +\sum_{n=1}^{l}\frac{q^{(A)}_{n,l}}{w_0^{2n}}{\cal
 I}^{(A)}_{n}\right)
  \Bigg] ,
   \label{gawa}
\eea
with 
\bea
{\cal I}^{(A)}_{n}
\!\!\!\!&=&\!\!\!\!%
(n+1)w^{2n+1}\cj_{1}-w^{2n+2}\cj_{0} ,
\\
q^{(A)}_{l,l}
\!\!\!\!&=&\!\!\!\!%
-{l\over (2l+1)(l+1)} d_{2l} ,
\\
q^{(A)}_{n,l}
\!\!\!\!&=&\!\!\!\!%
(-2)^{l-n}{2l+1\over 2n+1}
 \left[ \prod_{k=n+1}^{l}{k(k-1)(k+1)\over 2k+1}\right] 
 q^{(A)}_{l,l} , \quad \textrm{for}~~l>n\geq 1.
\eea

\end{appendix}

\newpage

\end{document}